\documentclass[11pt]{article}

\usepackage{graphicx}
\usepackage{amsmath}
\usepackage{amsfonts}
\usepackage{amssymb}
\usepackage{verbatim}

\renewcommand{\arraystretch}{1.3}

\catcode`\@=11
\def\marginnote#1{}

\newcount\hour
\newcount\minute
\newtoks\amorpm
\hour=\time\divide\hour by60
\minute=\time{\multiply\hour by60 \global\advance\minute by-\hour}
\edef\standardtime{{\ifnum\hour<12 \global\amorpm={am}%
        \else\global\amorpm={pm}\advance\hour by-12 \fi
        \ifnum\hour=0 \hour=12 \fi
        \number\hour:\ifnum\minute<10 0\fi\number\minute\the\amorpm}}
\edef\militarytime{\number\hour:\ifnum\minute<10 0\fi\number\minute}

\def\draftlabel#1{{\@bsphack\if@filesw {\let\thepage\relax
      \xdef\@gtempa{\write\@auxout{\string
          \newlabel{#1}{{\@currentlabel}{\thepage}}}}}\@gtempa \if@nobreak
    \ifvmode\nobreak\fi\fi\fi\@esphack} \gdef\@eqnlabel{#1}}
    \def\@eqnlabel{}
\def\@vacuum{}
\def\draftmarginnote#1{\marginpar{\raggedright\scriptsize\tt#1}}

\def\draft{
  \oddsidemargin -.5truein
  \def\@oddfoot{\footnotesize \sl preliminary draft \hfil
    \rm\thepage\hfil\sl\today\quad\militarytime}
  \let\@evenfoot\@oddfoot \overfullrule 3pt
    \let\label=\draftlabel
    \let\marginnote=\draftmarginnote
  \def\@eqnnum{(\theequation)\rlap{\kern\marginparsep\tt\@eqnlabel}%
    \global\let\@eqnlabel\@vacuum}

  }

\makeatletter
\newdimen\normalarrayskip              
\newdimen\minarrayskip                 
\normalarrayskip\baselineskip
\minarrayskip\jot
\newif\ifold             \oldtrue            \def\new{\oldfalse}
\def\arraymode{\ifold\relax\else\displaystyle\fi} 
\def\eqnumphantom{\phantom{(\theequation)}}     
\def\@arrayskip{\ifold\baselineskip\z@\lineskip\z@
     \else
     \baselineskip\minarrayskip\lineskip2\minarrayskip\fi}
\def\@arrayclassz{\ifcase \@lastchclass \@acolampacol \or
\@ampacol \or \or \or \@addamp \or
   \@acolampacol \or \@firstampfalse \@acol \fi
\edef\@preamble{\@preamble
  \ifcase \@chnum
     \hfil$\relax\arraymode\@sharp$\hfil
     \or $\relax\arraymode\@sharp$\hfil
     \or \hfil$\relax\arraymode\@sharp$\fi}}
\def\@array[#1]#2{\setbox\@arstrutbox=\hbox{\vrule
     height\arraystretch \ht\strutbox
     depth\arraystretch \dp\strutbox
     width\z@}\@mkpream{#2}\edef\@preamble{\halign
\noexpand\@halignto
\bgroup \tabskip\z@ \@arstrut \@preamble \tabskip\z@ \cr}%
\let\@startpbox\@@startpbox \let\@endpbox\@@endpbox
  \if #1t\vtop \else \if#1b\vbox \else \vcenter \fi\fi
  \bgroup \let\par\relax
  \let\@sharp##\let\protect\relax
  \@arrayskip\@preamble}
\def\eqnarray{\stepcounter{equation}%
              \let\@currentlabel=\theequation
              \global\@eqnswtrue
              \global\@eqcnt\z@
              \tabskip\@centering
              \let\\=\@eqncr

 \halign to \displaywidth\bgroup
    \eqnumphantom\@eqnsel\hskip\@centering
    $\displaystyle \tabskip\z@ {##}$%
    \global\@eqcnt\@ne \hskip 2\arraycolsep
         $\displaystyle\arraymode{##}$\hfil
    \global\@eqcnt\tw@ \hskip 2\arraycolsep
         $\displaystyle\tabskip\z@{##}$\hfil
         \tabskip\@centering
    &{##}\tabskip\z@\cr}
\newfont{\hr}{msbm10}
\newfont{\ams}{msam10}

\hoffset= - 0.9in         

\newdimen\linethick  \linethick=0.4pt
\newdimen\hboxitspace    \hboxitspace=5pt
\newdimen\vboxitspace    \vboxitspace=5pt

\def\fr#1{%
\beq\new
\vcenter{
\hrule height\linethick
          \hbox{\vrule width\linethick
                \kern\hboxitspace
                \vbox{\kern\vboxitspace
                      \hbox{$\begin{array}{c}\displaystyle#1
         \end{array}$}%
                      \kern\vboxitspace}%
                \kern\hboxitspace
                \vrule width\linethick}%
          \hrule height\linethick}%
\eeq}

\def\beq{\begin{equation}}
\def\eeq{\end{equation}}
\def\ba{\beq\new\begin{array}{c}}
\def\ea{\end{array}\eeq}
\def\be{\ba}
\def\ee{\ea}

\def\F{{\cal F}}

\def\d{\partial}
\def\N2{${\cal N}=2$}

\def\1N{${\cal N}=1$}
\def\4N{${\cal N}=4$}
\def\nn{\nonumber}
\def\half{{\textstyle{1\over2}}}

\newcommand{\rf}[1]{(\ref{#1})}

\newdimen\linethick  \linethick=0.4pt
\newdimen\hboxitspace    \hboxitspace=5pt
\newdimen\vboxitspace    \vboxitspace=5pt

\def\fr#1{%
\beq\new
\vcenter{
\hrule height\linethick
          \hbox{\vrule width\linethick
                \kern\hboxitspace
                \vbox{\kern\vboxitspace
                      \hbox{$\begin{array}{c}\displaystyle#1
         \end{array}$}%
                      \kern\vboxitspace}%
                \kern\hboxitspace
                \vrule width\linethick}%
          \hrule height\linethick}%
\eeq}

\renewcommand{\d}{\partial}

\def\F{\mathcal{F}}

\renewcommand{\tt}[1][mer]{\hbox{\tiny{#1}}}

\title{{\bf
Zamolodchikov asymptotic formula
and instanton expansion in ${\cal N}=2$ SUSY
$N_f=2N_c$ QCD} \vspace{.5cm}}
\author{{\bf A. Marshakov}\thanks{E-mail: \ mars@itep.ru; mars@lpi.ru}\ ,\ \
\date{ } 
{\bf A. Mironov}\thanks{E-mail:
\ mironov@itep.ru; mironov@lpi.ru}
\date{ } \\
{\small {\it Theory Department, Lebedev Physics Institute}
and {\it ITEP, Moscow, Russia}}\\ \\
{\bf A.Morozov}\thanks{E-mail: \ morozov@itep.ru}
\date{ } \\ {\small
{\it ITEP, Moscow, Russia}}}

\begin{document}

%

\setcounter{footnote}{3}

\setcounter{tocdepth}{3}

\maketitle

\vspace{-9cm}

\begin{center}
\hfill FIAN/TD-20/09\\
\hfill ITEP/TH-42/09
\end{center}

\vspace{6cm}

\begin{abstract}
The AGT relations allow to convert the Zamolodchikov asymptotic
formula for conformal block into the instanton expansion of the
Seiberg-Witten prepotential for the theory with two colors and four
fundamental flavors. This provides an explicit formula for the
instanton corrections in this model. The answer is especially
elegant for vanishing matter masses, then the bare charge of gauge
theory $q_0 = e^{i\pi\tau_0}$ plays the role of a branch point on
the spectral elliptic curve. The exact prepotential at this point is
${\cal F} = {1\over 2\pi i}a^2\log q$ with $q\neq q_0$, unlike the
case of another conformal theory, with massless adjoint field.
Instead, $16q_0 = \theta_{10}^4/\theta_{00}^4(q) = 16q(1+O(q))$,
i.e. the Zamolodchikov asymptotic formula gives rise, in particular,
to an {\em exact} non-perturbative beta-function so that the
effective coupling differs from the bare charge by infinite number
of instantonic corrections.
\end{abstract}

\begin{center}
{\large {\it To the memory of Alesha Zamolodchikov}}
\end{center}

\section{Introduction}

The AGT relation \cite{AGT}-\cite{AGTlast},
originally motivated by consideration of $5$-brane compactifications
on Riemann surfaces \cite{brane},
expresses the $2d$ conformal blocks \cite{BPZ}-\cite{ZamAT}
through Nekrasov functions \cite{Nek}-\cite{Neklast},
which in the limit
$\epsilon_1,\epsilon_2\to 0$
reproduce \cite{NO,MN,Mquas} instanton contributions to the Seiberg-Witten
prepotentials \cite{SW1}-\cite{Khrev}, describing the
low-energy phases of multidimensional ${\cal N}=2$ SUSY gauge theories.
In the simplest case of the $4$-point Virasoro conformal
block, the relevant Seiberg-Witten model is $N_f=2N_c$ gauge theory in $4d$
with $N_c=2$, i.e. with two colors. This theory has vanishing beta-function,
and the asymptotically free models with less flavors arise when some masses
(and, therefore, some dimensions of external fields
in the conformal blocks) get large and decouple \cite{Gnc,mmm2}.

If instead one considers large intermediate dimension $\Delta$,
the situation gets even more interesting.
When translated by the AGT relation into
the language of SUSY gauge theory, the large $\Delta$
expansion turns into the instanton expansion
(in $\Delta^{-1} \sim 1/a^2$).
More accurately, the number of instantons is controlled
by power of the double-ratio $x$-parameter in the conformal block,
to be identified, up to a numeric factor, with the bare coupling
$q_0 = e^{i\pi\tau_0} = e^{i\vartheta_0-8\pi^2/g_0^2}$.
At the {\it conformal} point\footnote{
We emphasize that here
``{\it conformal}'' (in italic) refers to conformal invariance in $4d$,
while the same word in ``the conformal block'' refers to $2d$
conformal invariance.}
with all vanishing masses of matter hypermultiplets, the prepotential is just
${\cal F} = \half \tau a^2$ with no $1/a^2$-corrections, just as follows from dimensional
reasoning.
However, this does not mean that instanton
corrections are absent! In fact, the effective coupling $\tau \neq \tau_0$
{\it differs} from the bare charge,
which is an exact and explicit example of topological charge renormalization
\cite{KM}-\cite{Ape}.
Finding an exact form for the function $\tau=\tau(\tau_0)$
or, better, $q=q(q_0)$, is one of the long-standing problems in
Seiberg-Witten theory (see e.g. \cite{SW2,AS,Kh,Min}).

Naively one may think that $\tau=\tau_0$, i.e. is not renormalized at all
for vanishing masses in $N_f=2N_c$ model, like it happens in the theory
with massless adjoint supermultiplet \cite{MontOl,DW,IM,Mb,GorMir}.
The approach to Seiberg-Witten theory, based on spectral curves
and integrable systems \cite{GKMMM,HO,AS,GMMM} leaves this issue aside,
since only the {\it effective}
coupling $\tau$ is derived as the period matrix of the Seiberg-Witten curve.
However, a direct instanton calculus (see e.g. \cite{Kh,Nek,Sha})
immediately demonstrates that the dependence $\tau=\tau(\tau_0)$
on the bare charge $\tau_0$ is nontrivial.
This can be seen directly by extrapolating the Nekrasov
functions from $N_f<2N_c$ case to $N_f=2N_c$:
the result is obviously non-trivial (the instantonic contributions contain ratios
$a_i^{N_f}/\prod_{j\neq i} a_{ij}$, clearly different from zero for non-vanishing values
of the condensates).
Still, this does not give immediately the dependence $q(q_0)$, which can be
possibly extracted by a careful treatment of the quasiclassical regime
along the lines of \cite{NO,MN,Mquas}.

However, after discovery of the AGT relations, at least, for $N_c=2$
and $N_f=2N_c=4$, one can simply {\it read} the answer for
this dependence from
the wonderful papers by Alesha Zamolodchikov \cite{Zam}, where the large-$\Delta$
asymptotics of the conformal block has been found.
The leading behavior for the conformal block (exact
at the {\it conformal point} and for $c=1$) is just
\be
{\cal B}_\Delta(x) \sim (16q)^\Delta
\ee
with\footnote{
Remarkably, this dressing formula
reproduces the geometrical engineering prediction
of ref.\cite{Mar,AGT}.
We are indebted to Marcos Marino for
attracting our attention to this important
paper.
}
\be
\boxed{
x= 16q_0 = \frac{\theta^4_{10}(q)}{\theta_{00}^4(q)}
= 16q\prod_{n=1}^\infty \left(\frac{1+q^{2n}}{1+q^{2n-1}}\right)^8
}
\label{xvsq}
\ee
where the Jacobi theta-constants are
\be
\theta_{00}(q) = 1 + 2\sum_{n=1}^\infty q^{n^2}
= \prod_{n=1}^\infty (1-q^{2n})(1+q^{2n-1})^2,  \\
\theta_{10}(q) = 2q^{1/4}\sum_{n=0}^\infty q^{n(n+1)}
= 2q^{1/4}\prod_{n=1}^\infty (1-q^{2n})(1+q^{2n})^2
\label{theta}
\ee
It means that the prepotential in the $N_f=2N_c=4$ Seiberg-Witten theory
with vanishing masses is
\be
{\cal F}(a) = {1\over 2\pi i}a^2\log q = \half a^2\tau
\label{preco}
\ee
which depends nontrivially over the bare coupling, in contrast
to the classical prepotential
\be
{\cal F}_{cl}(a) = {1\over 2\pi i}a^2\log q_0 = \half a^2\tau_0
\label{precl}
\ee
due to \rf{xvsq}. The classical prepotential \rf{precl} does not
get corrections in perturbation theory, since the beta-function vanishes
for $N_f=2N_c$, but is nontrivially renormalized non-perturbatively.

The spectral curve for the $N_c=2$, $N_f=4$ Seiberg-Witten theory
can be written as \cite{HO,GMMM}
\be
Y^2 = (X^2-u)^2 -  Q_4(X|\mu)
\label{specm}
\ee
with the polynomial $Q_4(X|\mu;\tau)$ of degree $4$ which depends on masses.
It is endowed
with the meromorphic generating differential
\be
dS \sim x\left({2XdX\over Y}-{X^2-u\over 2Y}{dQ_4(X|\mu)\over Q_4(X|\mu)}\right)
\ee
with residues proportional to the fundamental masses. The prepotential is defined
by $a^D = {\d\F\over\d a}$, where
\be
\label{swper}
a = \oint_A dS,\ \ \ a^D = \oint_B dS
\ee
At the {\it conformal} point this turns into
\be
Y^2 = (X^2-u)^2 -  \zeta X^4
\\
dS \sim u{dX\over Y}
\label{specp}
\ee
so that the Seiberg-Witten form becomes {\it holomorphic},
while $\zeta=\zeta(\tau)$ is a dimensionless parameter, which can
depend therefore only upon the effective coupling. One gets
hence from \rf{swper}, \rf{specp} that at the {\it conformal}
point
\be
\label{adac}
a^D = \tau a
\ee
where $\tau$ is the modulus (period matrix) of \rf{specm} so that
the prepotential is indeed given by \rf{preco}.

The curve \rf{specp} can be re-written in the form
\be
\eta^2 = \xi(\xi-1)(\xi-x)
\label{swzam}
\ee
related to \rf{specp} by a fractional-linear transformation
with\footnote{Sometimes a different parametrization
\be
{\tilde\zeta} = \frac{4x(x-1)}{(2x-1)^2} = -\frac{4\theta_{10}^4\theta_{01}^4}
{(\theta_{01}^4-\theta_{10}^4)^2}
\ee
is chosen, which is related to \rf{zx} by modular transformation
${\tilde\zeta}(x)=\zeta(1-x^{-1})$.}
\be
\zeta = {4x\over (1+x)^2} = \frac{4\theta_{00}^4\theta_{10}^4}
{(\theta_{00}^4+\theta_{10}^4)^2}
\label{zx}
\ee
Already the form of the elliptic curve (\ref{swzam}) leads
immediately to the relation (\ref{xvsq}). Indeed (see e.g.
\cite{BE}), one can associate with this curve the following values of the
Weierstrass $\wp$-function at half-periods: $e_1={2-x\over 3}$,
$e_2={2x-1\over 3}$ and $e_3=-{1+x\over 3}$ . Then,
\be
x={e_2-e_3\over
e_1-e_3}=\kappa^2=\frac{\theta^4_{10}(q)}{\theta_{00}^4(q)}
= 16q\prod_{n=1}^\infty \left(\frac{1+q^{2n}}{1+q^{2n-1}}\right)^8
\ee
In the weak coupling limit,
when $x\to 0$, $q\to 0$ and, therefore, $x=16q_0$.

We see that in the parameterizations \rf{swzam} the only nontrivial
branching point coincides with the bare coupling \rf{xvsq}! The bare
coupling therefore does not transform in a right way under the duality
transformation and this is a geometric reason for the nontrivial
instantonic renormalization \rf{xvsq}. It deserves noting that
the curve \rf{swzam} has appeared in the study \cite{Zam,ZamAT} of the
quasiclassical limit of conformal block.

Eq.(\ref{xvsq}) is the topological-charge renormalization
formula of our main interest. At weak coupling the
$q=e^{i\pi\tau}$-parameter of the torus
is close to $q_0={x\over 16}$, but gets RG-dressed
in this wonderful algebro-geometric way
(quite similar to the {\it dream-like} behavior in other
applications \cite{Pru}).

Note, however, that extracting instanton expansion of the
prepotential from the exact algebro-geometric solution is
a nontrivial technical problem (see e.g. \cite{Ph,Mas,MN}).
In the rest of the text we provide more technical
details and illustrations about the large-$\Delta$ asymptotics
of the AGT relation, using directly the Zamolodchikov asymptotic expansion.
We use notation and formulas from \cite{MMMagt}.

\newpage

\section{Asymptotic formula for the conformal block \cite{Zam}}

\unitlength 1mm 
\linethickness{0.4pt}
\ifx\plotpoint\undefined\newsavebox{\plotpoint}\fi 
\begin{picture}(47.94,-06.98)(-115,25)
\put(18.061,18.879){\line(1,0){22.818}}
\multiput(8.993,24.899)(.049970214,-.033593421){177}{\line(1,0){.049970214}}
\multiput(17.838,18.953)(-.058757277,-.03364733){148}{\line(-1,0){.058757277}}
\multiput(47.94,24.899)(-.039031239,-.033633302){179}{\line(-1,0){.039031239}}
\multiput(40.953,18.879)(.047206701,-.03364733){148}{\line(1,0){.047206701}}
\put(9.96,27){$\Delta_1$}
\put(9.96,10){$\Delta_2$}
\put(47.048,27){$\Delta_3$}
\put(47.048,10){$\Delta_4$}
\put(28.392,20.662){$\Delta$}
\end{picture}

\noindent
By definition, the $4$-point Virasoro conformal block is
a formal series
\be
\hspace{-5cm}{\cal B}^{\Delta}_{\Delta_1\Delta_2;\Delta_3\Delta_4}(x)
= \sum_{|Y|=|Y'|} x^{|Y|}
\gamma_{\Delta\Delta_1\Delta_2}(Y)Q^{-1}_\Delta(Y,Y')
\gamma_{\Delta\Delta_3\Delta_4}(Y') =
\\
= 1 + x\frac{D_{12}^{(1)}D_{34}^{(1)}}{2\Delta}
+ x^2\Big(D_{12}^{(2)},\ D_{12}^{(1)}(D_{12}^{(1)}+1)\Big)
\left( \begin{array}{cc} \frac{c}{2} + 4\Delta & 6\Delta \\
6\Delta & 4\Delta(2\Delta+1)\end{array}\right)^{-1}
\left(\begin{array}{c}
D_{34}^{(2)}\\ D_{34}^{(1)}(D_{34}^{(1)}+1)
\end{array}\right)
+ \ldots
\label{cbser}
\ee
where we have introduced a condensed notation $D_{ij}^{(k)} =
\Delta + k\Delta_i - \Delta_j$.
In fact, this formal series is a nice analytic function
of its arguments, possessing well-defined singularities
and satisfying certain equations \cite{BPZ,ZZ,Zam}. In particular,
according to \cite{Zam}, the conformal block \rf{cbser}
can be also represented as
\be
\boxed{
{\cal B}^{\Delta}_{\Delta_1\Delta_2;\Delta_3\Delta_4}(x) =
\left({16q\over x}\right)^{\Delta - \frac{c-1}{24}} (1-x)^{\frac{c-1}{24}-\Delta_1-\Delta_3}
\theta_{00}(q)^{\frac{c-1}{2}-4(\Delta_1+\Delta_2+\Delta_3+\Delta_4)}
H^{\Delta}_{\Delta_1\Delta_2;\Delta_3\Delta_4}(x),
}
\label{Zaf}
\ee
where
$H^{\Delta}_{\Delta_1\Delta_2;\Delta_3\Delta_4}(x) = 1 + \sum_{k>0}h_kq^k =
1 + O(\Delta^{-1})$ is equal to unity at $\Delta\rightarrow \infty$.
This formula can be considered as a partial summation of the series
(\ref{cbser}), and $q$ in (\ref{Zaf}) is related to $x$ by
eq.(\ref{xvsq}).
The function $H^{\Delta}_{\Delta_1\Delta_2;\Delta_3\Delta_4}$
behaves nicely if considered as a function of $q$
rather than $x$, still it is quite complicated and not well understood yet,
for the latest results about this function see \cite{FL}.
Here we only need its asymptotic behavior at large $\Delta$.

In the rest of this section we present some illustration of the Zamolodchikov
formula (\ref{Zaf}),
demonstrating that it is indeed a very {\it explicit} statement.
It has been proved long ago in \cite{Zam}, and we do not discuss
the original proof in this letter.

\subsection{Explicit expansions}

Making use of \rf{theta}, one can invert (\ref{xvsq})
\be
16 q = x\left(1 + \frac{1}{2}x +\frac{21}{64}x^2 + \frac{31}{128}x^3 +
\frac{6257}{32768}x^4 + \frac{10293}{65536}x^5
+ \ldots\right)
\label{qthrx}
\ee
and represent in a similar form the other entries of (\ref{Zaf}): the factor
\be
\left(\frac{16q}{x}\right)^\Delta =
1+\frac{\Delta}{2}x+\left(\frac{21\Delta}{64}
+\frac{\Delta(\Delta-1)}{8}\right)x^2 +
\\
+\left(\frac{31\Delta}{128} + \frac{21\Delta(\Delta-1)}{128}
+\frac{\Delta(\Delta-1)(\Delta-2)}{48}\right)x^3 +   \\
+\left(\frac{6257\Delta}{32768} + \frac{1433\Delta(\Delta-1)}{8192}
+\frac{21\Delta(\Delta-1)(\Delta-2)}{512}
+\frac{\Delta(\Delta-1)(\Delta-2)(\Delta-3)}{384}\right)x^4
+\ldots
\label{qxDel}
\ee
and the theta-constant
\be
\theta_{00}(q) =
1+\frac{x}{8}+\frac{x^2}{16}+\frac{21x^3}{512}+\frac{993x^4}{32768}
+\frac{6273x^5}{262144}+O(x^6)
\label{thex}
\ee
For their logarithms one gets
\be
\log\left(\frac{16q}{x}\right) = \frac{x}{2}+\frac{13x^2}{64}
+\frac{23x^3}{192}+\frac{2701x^4}{32768}+\frac{5057x^5}{81920}+O(x^6)
\ee
\be
\log\theta_{00}(q) = \frac{x}{8}+\frac{7x^2}{128}
+\frac{13x^3}{384}+\frac{791x^4}{32768}+\frac{1523x^5}{81920}+O(x^6)
\ee

\subsection{(\ref{cbser}) vs (\ref{Zaf}): the first order in $x$}

Making use of the first terms in (\ref{qxDel}) and (\ref{thex}),
one gets from (\ref{Zaf}) in the first order in $x$:
\be
1 + x \left\{\frac{1}{2}\left(\Delta-\frac{c-1}{24}\right) +
\left(\Delta_1+\Delta_3 - \frac{c-1}{24}\right) + \right.
\\
\left.+\frac{1}{8}
\left(\frac{c-1}{2} - 4(\Delta_1+\Delta_2+\Delta_3+\Delta_4)\right)\right\}
(1 + xh_1 + \ldots) =
\ee
\vspace{-0.3cm}
\be
= 1 + \frac{x}{2}\Big(\Delta+\Delta_1-\Delta_2+\Delta_3-\Delta_4+2h_1\Big)
+ \ldots
\ee
Comparing this with the first-order term in (\ref{cbser}), one obtains that
\be
h_1 = \frac{(\Delta_1-\Delta_2)(\Delta_3-\Delta_4)}{2\Delta}
\label{h1}
\ee
and, indeed, $h_1 = O(\Delta^{-1})$.

\subsection{(\ref{cbser}) vs (\ref{Zaf}): the second order in $x$
at {\it conformal point}}

In the second order in $x$ formulas get more complicated,
we begin with the simplified case, where all four external dimensions
vanish $\Delta_{1,2,3,4}=0$.
Then,  the  series (\ref{cbser}) becomes
\be
\label{cpcb}
1 + x\frac{\Delta}{2} + x^2\Big(\Delta,
\Delta(\Delta+1)\Big)
\left( \begin{array}{cc} \frac{c}{2} + 4\Delta & 6\Delta \\
6\Delta & 4\Delta(2\Delta+1)\end{array}\right)^{-1}
\left(\begin{array}{c}
\Delta\\ \Delta(\Delta+1)\end{array}\right) + O(x^3) =
\\
= 1 + \frac{x\Delta}{2} +
\frac{x^2\Delta^2}{8}\cdot\frac{\Delta^2 + \left(4+\frac{c}{2}\right)\Delta
+ (c-4) + \frac{c}{2\Delta}}{\Delta^2 + \frac{c-5}{8}\Delta + \frac{c}{16}}
 + O(x^3) =
 \\
= 1 + \frac{x\Delta}{2} +
\frac{x^2}{8}\left(\Delta^2 + \frac{13\Delta}{8} +
\frac{1-c}{64} + O(\Delta^{-1})\right)
 + O(x^3)
\ee
Note that in these orders in $x$
the $c$-dependence appears only in the $O(\Delta^{-1})$ terms.
For vanishing external dimensions (\ref{Zaf}) reduces to
\be
\overbrace{\left(1+\frac{\Delta-\sigma}{2}x
+\left(\frac{13(\Delta-\sigma)}{64}
+\frac{(\Delta-\sigma)^2}{8}\right)x^2 + \ldots\right)
}^{(16q/x)^{\Delta-\sigma}}
\overbrace{
\left(1-x\sigma + \frac{\sigma(\sigma-1)}{2}x^2 + \ldots\right)
}^{(1-x)^\sigma} \cdot \\
\cdot\underbrace{\left(1+\frac{12\sigma}{8}x
+\left(\frac{12\sigma}{16}+\frac{12\sigma(12\sigma-1)}{2\cdot 8^2}\right)x^2
+\ldots\right)}_{\theta_{00}(q)^{12\sigma}}
\Big(1+h_2x^2+\ldots\Big) = \\ =
1 + \frac{x\Delta}{2} +
\frac{x^2}{8}\left(\Delta^2 + \frac{13\Delta}{8}
-\frac{3\sigma}{8} +h_2\right) + \ldots
\label{cpZa}
\ee
where $\sigma = \frac{c-1}{24}$ and we used the already evaluated
(\ref{h1}) implying that $h_1=0$ at the {\it conformal point}.
Comparing (\ref{cpZa}) with (\ref{cpcb}) one observes that, indeed,
$h_2 = O(\Delta^{-1})$.

\subsection{(\ref{cbser}) vs (\ref{Zaf}): the second order in $x$}

Restoring dependence on the external dimensions, one obtains instead of
(\ref{cpcb}) and (\ref{cpZa}):
\be
1 + x \frac{(\Delta+\Delta_1-\Delta_2)(\Delta+\Delta_3-\Delta_4)}{2\Delta}
+  \\
+ x^2\Big(\Delta+2\Delta_1-\Delta_2,\
(\Delta+\Delta_1-\Delta_2)(\Delta+\Delta_1-\Delta_2+1)\Big)\cdot
\\
\cdot
\left( \begin{array}{cc} \frac{c}{2} + 4\Delta & 6\Delta \\
6\Delta & 4\Delta(2\Delta+1)\end{array}\right)^{-1}
\left(\begin{array}{c}
\Delta+2\Delta_3-\Delta_4\\
(\Delta+\Delta_3-\Delta_4)(\Delta+\Delta_3-\Delta_4+1)
\end{array}\right) + \ldots = \nn
\ee
\be
= 1 + x \frac{(\Delta+\Delta_1-\Delta_2)(\Delta+\Delta_3-\Delta_4}{2\Delta}
+ \frac{x^2}{8}\left\{\Delta^2
+ \left(\frac{13}{8}+2(\Delta_1-\Delta_2+\Delta_3-\Delta_4)\right)\Delta
+\right. \\ \left.+
\left(\frac{1-c}{64} + \frac{9(\Delta_1+\Delta_3)-7(\Delta_2+\Delta_4)}{4}
+ (\Delta_1-\Delta_2)^2 + 4(\Delta_1-\Delta_2)(\Delta_3-\Delta_4)
+ (\Delta_3-\Delta_4)^2\right) + \right.  \\ +
\left(\frac{(c+1)(c+5)}{512}
- \frac{(c+3)(\Delta_1+\Delta_2+\Delta_3+\Delta_4)}{32}+ \right.
\\
+ \frac{1}{8}\Big((\Delta_1-\Delta_2)^2+(\Delta_3-\Delta_4)^2
+ 20(\Delta_1\Delta_3+\Delta_2\Delta_4)
- 12(\Delta_1\Delta_4+\Delta_2\Delta_3)\Big)
+ \\ \left.\left.
+ 16(\Delta_1-\Delta_2)(\Delta_3-\Delta_4)
(\Delta_1-\Delta_2+\Delta_3-\Delta_4)
\right)\frac{1}{\Delta}
+ O(\Delta^{-2})
\right\} + \ldots =\nn
\ee
\be
=
\left\{1 + \frac{x}{2}(\Delta + \Delta_1-\Delta_2+\Delta_3-\Delta_4)
+ \frac{x^2}{8}
\left[\Delta^2
+ \left(\frac{13}{8}+2(\Delta_1-\Delta_2+\Delta_3-\Delta_4)\right)\Delta
+\right.\right. \\ \left.\left.+
\left(\frac{1-c}{64} + \frac{9(\Delta_1+\Delta_3)-7(\Delta_2+\Delta_4)}{4}
+ (\Delta_1-\Delta_2+\Delta_3-\Delta_4)^2\right)\right]
+ \ldots
\right\}\cdot \\\cdot
\left\{1 + x\frac{(\Delta_1-\Delta_2)(\Delta_3-\Delta_4)}{2\Delta}
+ \frac{x^2}{8\Delta}\left[\frac{(c+5)(c+1)}{512}
- \frac{(c+3)(\Delta_1+\Delta_2+\Delta_3+\Delta_4)}{32}
+ \right.\right. \\ \left.\left. +
\frac{1}{8}\Big((\Delta_1-\Delta_2)^2+(\Delta_3-\Delta_4)^2
+ 20(\Delta_1\Delta_3+\Delta_2\Delta_4)
- 12(\Delta_1\Delta_4+\Delta_2\Delta_3)\Big)
+ O(\Delta^{-1})\right] + \ldots\right\}
\label{Zaf2}
\ee
The l.h.s. of this equality represents the original expansion of the
conformal block (\ref{cbser}), while
the r.h.s. represents the expansion of the Zamolodchikov formula (\ref{Zaf}).
This ends our demonstrations of how eq. (\ref{Zaf}) works, which can be easily
extended to higher orders with the help of computer simulations, and in the next
section we are going directly to the consequences of the Zamolodchikov formula
for the Seiberg-Witten prepotential.

\section{Implication for the SW prepotential}

The Nekrasov functions appear to be a clever regularization
of the instantonic sums in multidimensional supersymmetric gauge theories,
and originally auxiliary $\epsilon$-parameters turn out to be
the crucial modification. In particular, the AGT relation implies that the $2d$ central
charge is
$c = 1 + \frac{6(\epsilon_1+\epsilon_2)^2}{\epsilon_1\epsilon_2}$, so that the
$\epsilon$-parameters have a clear sense from the two-dimensional point of view.
In order to get the Seiberg-Witten prepotentials, one should consider
the limit, when both $\epsilon_1$ and $\epsilon_2$ go to zero,
and extract the most singular term
\be
{\cal F}_{inst} = {\cal F} - {\cal F}_{cl} =
\lim_{\epsilon_1,\epsilon_2\to 0}
{-\epsilon_1\epsilon_2\over 2\pi i}
\log\Big(
(1-x)^{-2\alpha_1\alpha_3/\epsilon_1\epsilon_2}
{\cal B}^\Delta_{\Delta_1\Delta_2;\Delta_3\Delta_4}(x)\Big)
\label{FvsB}
\ee
where the
parameter $x$ is associated with the bare charge
$x = 16q_0 = 16e^{i\pi\tau_0}$ and the classical prepotential \rf{precl},
while the external dimensions
$\Delta_I$ are related to the fundamental masses $\mu_I$, $I=1,\ldots,4$.
If one puts $c=1$, the relation is especially simple:
$\Delta_I = \frac{\alpha_I^2}{-\epsilon_1\epsilon_2}$ and
\be
\mu_{1,2} = \alpha_1\pm \alpha_2,\ \ \ \ \mu_{3,4} = \alpha_3\pm\alpha_4
\label{muagt}
\ee
The same parameters $\alpha_1$ and $\alpha_3$ appear in
the additional $U(1)$-factor
$(1-x)^{-2\alpha_1\alpha_3/\epsilon_1\epsilon_2}$ at the r.h.s. of
(\ref{FvsB}), this factor plays a crucial role in restoring the
symmetry between the four masses $\mu_I$ in the prepotential.
The conformal block (\ref{cbser}) is {\it not} symmetric
in the external dimensions, since the vertex
operators are located at different points on the Riemann sphere.

After that, the expansion of the asymptotic formula (\ref{Zaf})
considerably simplifies, since all the terms with $\sigma = \frac{c-1}{24}$
can be omitted, and only the leading
powers of dimensions should be kept: the corrections vanish in the limit
of small $\epsilon_{1,2}$-parameters. It is still not obvious how
does the prepotential look like, and even why it is
symmetric in $\mu_{1,2,3,4}$. Therefore, we start with an illustration,
explicitly evaluating the first two orders in $x$.

\subsection{Expansion of the prepotential from the conformal block}

Taking logarithm of the r.h.s. of (\ref{Zaf2}) one obtains:
\be
\log\,{\cal B}(x) =
\frac{x}{2}(\Delta + \Delta_1 - \Delta_2+\Delta_3-\Delta_4)
+ \frac{x^2}{8}\left(\frac{13\Delta}{8}
+ \frac{9(\Delta_1+\Delta_3) - 7(\Delta_2+\Delta_4)}{4}\right) + \ldots
   \\
\ldots + x\frac{(\Delta_1-\Delta_2)(\Delta_3-\Delta_4)}{2\Delta} +
\\
+ \frac{x^2}{64\Delta}\Big(
(\Delta_1-\Delta_2)^2 + (\Delta_3-\Delta_4)^2 +
20(\Delta_1\Delta_3+\Delta_2\Delta_4)
- 12(\Delta_1\Delta_4+\Delta_2\Delta_3) +
  \\ \left. +
\frac{3}{16} -
({\Delta_1+\Delta_2+\Delta_3+\Delta_4}) +  O(\Delta^{-1})\right) + \ldots
\label{logB}
\ee
The first line here comes from the logarithm of the Zamolodchikov asymptotics, i.e.
from the divergent at $\Delta\to\infty$ part of \rf{Zaf};
note that the $\Delta^2$-terms growing  as $(\epsilon_1\epsilon_2)^{-2}$ in the limit
$\epsilon_{1,2}\rightarrow 0$ have disappeared after taking logarithm of \rf{Zaf2}.
The other three lines in \rf{logB} present the contribution of the correction
from $\log H=O(\Delta^{-1})$. The terms, which survive in this limit, are contained in the
second and third lines, they are all of the form ${\Delta_I\Delta_J\over\Delta}$, and
the linear terms from the forth line disappear in the limit.

The first two lines in (\ref{logB}) are
not symmetric in four masses $\mu_{1,2,3,4}$, which cannot be
true for the prepotential: already in the linear order in $x$,
according to (\ref{muagt}), the term with
$(\Delta_1-\Delta_2)(\Delta_3-\Delta_4) = \mu_1\mu_2\mu_3\mu_4$
is symmetric, while
$\Delta_1-\Delta_2+\Delta_3-\Delta_4 =
\mu_1\mu_2 + \mu_3\mu_4$ is not.
Restoring the symmetricity comes from the $U(1)$-factor in the r.h.s. of \rf{FvsB}:
\be
-2\alpha_1\alpha_3\log(1-x) =
\frac{1}{2}(\mu_1+\mu_2)(\mu_3+\mu_4)
\left(x - \frac{x^2}{2} + \ldots\right)
\ee
which converts the $\mu^2$-term in the order $x$ into the symmetric
combination $\frac{x}{2}\sum_{i<j}\mu_i\mu_j$.

Analyzing in the same way the $x^2$-contributions, one gets
\be
2\pi i{\cal F}_{inst} = a^2
\log\left(\frac{16q}{x}\right) +
\Big(\epsilon_1\epsilon_2(\Delta_1+\Delta_3)-2\alpha_1\alpha_3\Big)
\log(1-x) +
\\
+ \epsilon_1\epsilon_2(\Delta_1+\Delta_2+\Delta_3+\Delta_4)
\log \left(\theta_{00}^4(q)\right)-
 \lim_{\epsilon_1\epsilon_2\rightarrow 0}
\epsilon_1\epsilon_2 \left(xh_1 + x^2\left(h_2-\frac{h_1^2}{2}\right)
+ \ldots\right) =   \\
= x\left(\frac{a^2}{2} + \alpha_1^2+\alpha_3^2 + 2\alpha_1\alpha_3
-\frac{1}{2}(\alpha_1^2+\alpha_2^2+\alpha_3^2+\alpha_4^2)
- \frac{(\alpha_1^2-\alpha_2^2)(\alpha_3^2-\alpha_4^2)}{2a^2}
\right) +\\
+ \frac{x^2}{8}\left(\frac{13a^2}{8} + \frac{1}{4}
(9\alpha_1^2-7\alpha_2^2+9\alpha_3^2-7\alpha_4^2 +32\alpha_1\alpha_3)
- \right.  \\ \left.
- \frac{(\alpha_1^2-\alpha_2^2)^2+(\alpha_3^2-\alpha_4^2)^2
+20(\alpha_1^2\alpha_3^2+\alpha_2^2\alpha_4^2)
-12(\alpha_1^2\alpha_4^2+\alpha_2^2\alpha_3^2)}{8a^2} + O(a^{-4})
\right) + O(x^3) =   \nn
\ee
\be
= \frac{x}{2}\left(a^2 +
\underbrace{\mu_1\mu_2}_{\alpha_1^2-\alpha_2^2}
+\underbrace{\mu_3\mu_4}_{\alpha_3^2-\alpha_4^2} +
\underbrace{(\mu_1+\mu_2)}_{2\alpha_1}\underbrace{(\mu_3+\mu_4)}_{2\alpha_3}
- \frac{\mu_1\mu_2\mu_3\mu_4}{2a^2}\right) +   \\ + \frac{x^2}{8}
\left(\frac{13a^2}{8} +  \frac{9(\mu_1+\mu_2)^2 - 7(\mu_1-\mu_2)^2
+ 9(\mu_3+\mu_4)^2 - 7(\mu_3-\mu_4)^2 + 32(\mu_1+\mu_2)(\mu_3+\mu_4)}{16}
- \right.  \\
- \frac{\mu_1^2\mu_2^2 + \mu_3^2\mu_4^2 +
\frac{5}{4}\Big((\mu_1+\mu_2)^2(\mu_3+\mu_4)^2
+ (\mu_1-\mu_2)^2(\mu_3-\mu_4)^2\Big)}{8a^2} +\\+\frac{
\frac{3}{4}\Big((\mu_1+\mu_2)^2(\mu_3-\mu_4)^2
+ (\mu_1-\mu_2)^2(\mu_3+\mu_4)^2\Big)}{8a^2}  + O(a^{-4})  \Big)
+ O(x^3) = \nn
\ee
\fr{\label{Finst}
= \frac{x}{2}\left(a^2 + \sum_{i<j} \mu_i\mu_j
- \frac{\mu_1\mu_2\mu_3\mu_4}{2a^2}\right) + \\
+ \frac{x^2}{64}
\left(13a^2 + \sum_i \mu_i^2 + 16\sum_{i<j}\mu_i\mu_j
-\frac{1}{a^2}\sum_{i<j}\mu_i^2\mu_j^2 -
\frac{16\mu_1\mu_2\mu_3\mu_4}{a^2}
 + O(a^{-4})  \right)
+ O(x^3)
}
One has obtained, therefore, the instanton expansion of the prepotential
in the $SU(2)$ \N2 supersymmetric gauge theory with $N_f=4$ matter
hypermultiples directly by expansion of the Zamolodchikov asymptotic formula
\rf{Zaf}. We see now that the coefficients in front
of different structures in \rf{Finst} are dressed {\it differently} by the
instanton corrections. Using the algorithm of the previous section, such a
calculation can be easily continued in order to get higher-instanton corrections
to the Seiberg-Witten prepotential.

\subsection{Exact renormalization from Zamolodchikov formula}

The Zamolodchikov formula allows one also to find renormalization
of three structures {\it exactly} in all orders of $x$:
\be
\boxed{
2\pi i{\cal F}_{inst} =
a^2 \log\left(\frac{16q}{x}\right)
- \frac{1}{4}\left(\sum_{i=1}^4 \mu_i\right)^2 \log(1-x)
-\frac{1}{2}\sum_{i=1}^4 \mu_i^2
\log \left(\theta_{00}^4(q)\right)+ O(a^{-2})
}
\label{Finstexact}
\ee
just coming from the divergent at $\Delta\to\infty$ part of \rf{Zaf}.
Two of the structures are independent of $a$ and, therefore,
are not well controlled by the spectral curve and formulas \rf{swper}.
More important, one should remember that, from the point
of view of Seiberg-Witten theory, the AGT relation
and, therefore, the results (\ref{Finstexact}) and (\ref{Finst})
describe the instanton corrections computed by the Nekrasov algorithm \cite{Nek},
i.e. computed using the $\epsilon$-regularized moduli space of instantons.
In the case of $N_f=2N_c$ they are different from the corrections,
evaluated earlier with the help of singular moduli space (see pp. 190-204 of
\cite{Khrev} for discussion of this difference and references).
For example, (\ref{Finst}) contains the contribution of one instanton,
and generally the sectors with odd instantonic charges give nontrivial
terms into $\F_{inst}$.
Also, the two-instanton contribution in \rf{Finst} is different
from $\frac{7x^2a^2}{64\cdot 243}$ found in \cite{Kh}.

It would be interesting to find a polynomial $Q_4(X|\mu;\tau)$
in (\ref{specm}) for non-vanishing masses $\mu_{1,2,3,4}$,
which reproduces the answer (\ref{Finst}).

\section{Conclusion}

To conclude, in this letter we have studied the large-$\Delta$ asymptotics
of the AGT relation for the $4$-point conformal block.
This adds only a little to testing the AGT relations themselves,
but this is hardly the main point of interest now, since a lot of
evidence has been already collected.
More important, we proposed an application to the beautiful asymptotic
formula (\ref{Zaf}), found by Alesha Zamolodchikov many years ago \cite{Zam}.
The AGT relation converts this formula into the asymptotics of the
Nekrasov partition function, which turns itself into the Seiberg-Witten
prepotential for still mysterious $4d$
{\it conformal} theory with $N_f=2N_c$. The Zamolodchikov formula
is valid for the Virasoro conformal block, transferred by the AGT relation
into the simplest case of the non-Abelian \N2 supersymmetric gauge theory with
$N_c=2$, nevertheless the result is spectacular.
It confirms the expectation that the instanton corrections are
non-trivial, despite the theory is conformal invariant, thus
providing an example of purely non-perturbative renormalization group
in supersymmetric gauge theory (not only in the sector of the topological charge,
but also of the ordinary coupling constant).
Moreover, the renormalization group evolution is described here by the nice
algebro-geometric formula (\ref{xvsq}).
Further understanding of this result
and its generalization  to $N_c>2$
would be very interesting from many points of view.

\section*{Acknowledgements}

Our work was partly supported by Russian Federal Nuclear Energy
Agency and by the joint grants 09-02-90493-Ukr, 09-02-93105-CNRSL,
09-01-92440-CE. The work of A.Mar. was also supported by Russian
President's Grants of Support for the Scientific Schools
NSh-1615.2008.2, by the RFBR grant 08-01-00667, and by the Dynasty
Foundation. The work of A.Mir. was partly supported by the RFBR
grant 07-02-00878, while the work of A.Mor. by the RFBR grant
07-02-00645; the work of A.Mir. and A.Mor. was also supported by
Russian President's Grants of Support for the Scientific Schools
NSh-3035.2008.2 and by the joint program 09-02-91005-ANF. A.Mar.
would like also to thank the Yukawa Institute for Theoretical
Physics in Kyoto, where the work on the final version of this paper
has been completed, for the warm hospitality.

\end{document}